\documentclass[aps, twocolumn, prb,showpacs,amsmath,amssymb,superscriptaddress,letterpaper]{revtex4}
\usepackage{times}
\usepackage{amsfonts}
\usepackage{mathrsfs}
\usepackage{graphicx}
\usepackage{dcolumn}
\usepackage{bm}
\usepackage{color}

\usepackage[colorlinks,bookmarks=false,citecolor=blue,linkcolor=red,urlcolor=blue]{hyperref}
\def\be{\begin{equation}}       \def\ee{\end{equation}}
\def\bea{\begin{eqnarray}}      \def\eea{\end{eqnarray}}

\begin{document}
\title{Design strong anomalous Hall effect via spin canting in antiferromagnetic nodal line materials}

\author{Congcong Le}
\affiliation{Max Planck Institute for Chemical Physics of Solids, 01187 Dresden, Germany}


\author{Claudia Felser}
\affiliation{Max Planck Institute for Chemical Physics of Solids, 01187 Dresden, Germany}

\author{Yan Sun}\email{Corresponding: ysun@cpfs.mpg.de}
\affiliation{Max Planck Institute for Chemical Physics of Solids, 01187 Dresden, Germany}

\date{\today}

\begin{abstract}
The interplay between magnetism and topological electronic structure offers a large freedom
to design strong anomalous Hall effect (AHE) materials.
A nodal line from band inversion is a typical band structure to
generate strong AHE. Whereas, in most collinear antiferromagnets (AFMs),
the integration of Berry curvatures on Brillouin zone is forced to zero by the joint $TO$ symmetry,
where $T$ and $O$ are time reversal and a space group operation, respectively. Even with inverted band structures, such kind of AFM cannot have AHE. Therefore, so far, AFM nodal line band structures
constructed by spin degenerated bands didn't get much attentions in AHE materials.
In this work, we illustrate that such kind of band structure indeed provides a promising
starting point to generated strong local Berry curvature by perturbations and,
therefore, strong intrinsic AHE. In specific AFM compounds of $A$MnBi$_2$($A$=Ca and Yb) with inverted band structure, we found a strong AHE induced by a weak spin canting, and due to nodal line in the band structure the anomalous Hall conductivity
keeps growing as the canting angle increases. Since such spin-canting can be adjusted via doping experimentally, it provides another effective strategy to generate and manipulate strong AHE

\end{abstract}

\pacs{75.85.+t, 75.10.Hk, 71.70.Ej, 71.15.Mb}

\maketitle
\section{Introduction}

In recent years, materials with strong anomalous Hall effect (AHE)\cite{Nagaosa2010} have attracted extensive attentions in the field of materials science and condensed matter physics, which is connected to fundamental topological band structure and potential applications in electronic devices. Since the intrinsic AHE can be understood as the integral of Berry curvature in the momentum space\cite{Fang2003,Haldane2004,Xiao2010}, band structures with strong Berry curvature are desired to have strong AHE. Weyl points and nodal lines are two kinds of promising band structures, which can generate strong local Berry curvature. With such guiding principle, strong AHE materials with both large anomalous Hall conductivity (AHC) and anomalous angle (AHA) were observed in ferromagnetic Weyl semimetal (WSM) Co$_3$Sn$_2$S$_2$\cite{Liu2018,Wang2018,Xu2018}, nodal semimetal of layered ferromagnetic compound Fe$_3$GeTe$_2$\cite{Cheng2018,Kyoo2018} and Heusler compound Co$_2$Mn(Ga/Al)\cite{Guin2019,Pei2020,Kubler2016,Noky2019,Manna2018,Kaustuv2018} et. al.

~~~A crucial property of Berry curvature is odd under time reversal operator\cite{Xiao2010}, and hence AHE can only exist in magnetic system in linear response region. Due to broken time reversal symmetry $T$, the ferromagnet can host non-zero AHE. However, in the collinear antiferromagnets (AFM), though the time reversal is broken, there are two kinds of combined symmetries $TO$ between time reversal operation $T$ and a space group operation $O$, where $O$ is a fractional translation or inversion symmetry. Then, the combined symmetry $TO$ can also change the sign of Berry curvature, and results in the cancellation of Berry curvature when integrated over the full Brillouin zone (BZ). Therefore, AHE can only exist in a few situations without combined symmetries $TO$, such as non-collinear AFM in FeMn, NiS$_2$, cubic Mn$_3$(Ir/Pt) and hexagonal Mn$_3$(Ge/Sn)\cite{Shindou2001,Chen2014,Kubler2014,Nakatsuji2015,Nayak2016,Zhang2017,dos2020}, and collinear AFM Ti$_2$MnAl and RuO$_2$\cite{Shi2018,Feng}. On the other hand, AMF with combined symmetries $TO$ almost did not attract much attention in  AHE materials, even with nodal-line-like special band structures. Indeed, as we will show bellow, by perturbation, the nodal-line-like special band structures in the AMF materials with combined symmetries maybe generate strong AHE.

\begin{figure}
\centerline{\includegraphics[width=0.5\textwidth]{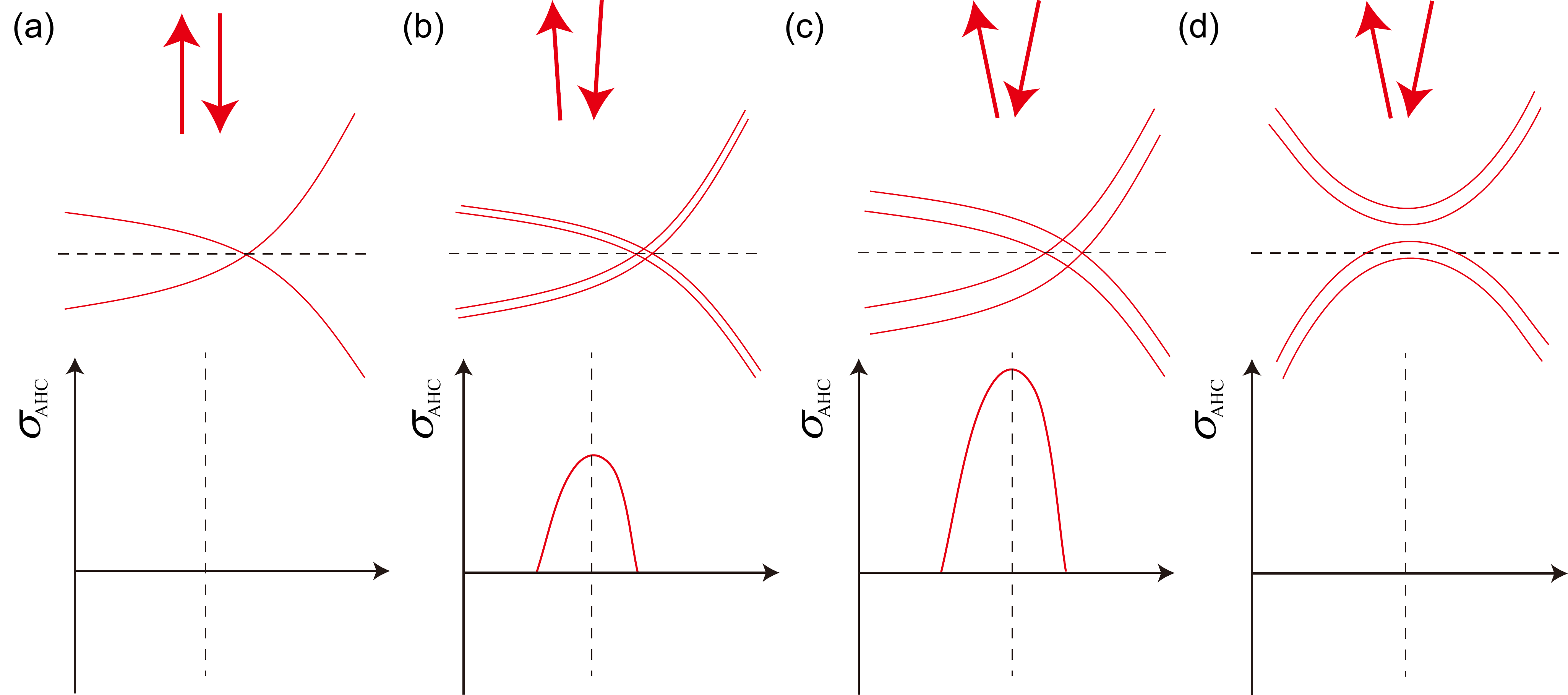}}
\caption{(color online) Schematic of AHE in the spin canted AFM order with(a-c) and without(d) band inversion, $\sigma_{AHC}$ is AHC and $E_f$ is Fermi level. (a) The canting angle $\theta$=0. (b) The small canting angle $\theta$. (c) The large canting angle $\theta$. (d) AHE in the spin canted AFM order without band inversion in the large canting angle $\theta$.
\label{fig0} }
\end{figure}

 ~~~In this work, we propose an effective strategy to generate strong Berry curvature by perturbations of spin-canting based on collinear AFM nodal line band structures with combined symmetries $TO$. Fig.\ref{fig0} shows schematic of AHE in the spin canted AFM order with inverted band
structure, where $\sigma_{AHC}$ is AHC. At the canting angle $\theta$=0, the bands are degeneracy due to time reversal and inversion symmetries, and hence $\sigma_{AHC}$ is zero, shown in Fig.\ref{fig0}(a). When the canting angle $\theta$ is small, the two degenerate bands are split because of broken time reversal symmetry, suggesting that $\sigma_{AHC}$ should be non-zero, as shown in Fig.\ref{fig0}(b). Moreover, since the band inversion already exists in spin-degenerated bands, a type of canting can lead to a shape change of Berry curvature and AHC. When the canting angle $\theta$ becomes large, the two degenerate bands are more split, and $\sigma_{AHC}$ maybe become larger, shown in Fig.\ref{fig0}(c). However, if the canted AFMs can not host topological band structures, and their AHEs are almost zero, shown in Fig.\ref{fig0}(d). Based on the above schematic, we perform an $ab~initio$ analysis for AHE as the canting angle changes in the spin canting C-type AFM materials AMnBi$_2$(A=Ca and Yb)\cite{He2012,Wang2016,Zhang2016,Qiu2018,Qiu2019,Sergey2019,Yang2020}, where the band inversion leads to a topological nodal ring without considering spin-orbital coupling (SOC). We find that the AHC always keeps growing as the canting angle increases, and doping electrons can reduce AHC while doping holes strengthen AHC.

\section{Method}\label{S1}

Our calculations are performed using density functional theory (DFT) as implemented in the Vienna ab initio simulation package (VASP) code \cite{Kresse1993,Kresse1996,Kresse1996B}. The Perdew-Burke-Ernzerhof (PBE) exchange-correlation functional and the projector-augmented-wave (PAW) approach are used. Throughout the work, the cutoff energy is set to be 550 eV for expanding the wave functions into a plane-wave basis. The Brillouin zone is sampled in the k space within the Monkhorst-Pack scheme\cite{Monkhorst}, and the k mesh used is $10 \times 10 \times 4$ on the basis of the equilibrium structure. In our calculations, a C-type AFM order along the c-axis, a spin-canting and SOC are included.

~~~To calculate AHC, we project the $ab~initio$ DFT Bloch
wave function into high symmetric atomic-orbital-like Wannier functions~\cite{Yates2007} with diagonal position operator, as performed in the Vienna ab initio simulation package (VASP) code \cite{Kresse1993,Kresse1996,Kresse1996B}. For obtaining precise Wannier functions, we include the outermost s- and d-orbital for Ca, d-orbital for element Mn and p-orbital for element Bi, which guarantees the full bands overlap from ab-initio and Wannier functions.

\section{Symmetry analysis }\label{S2}

\begin{figure}
\centerline{\includegraphics[width=0.5\textwidth]{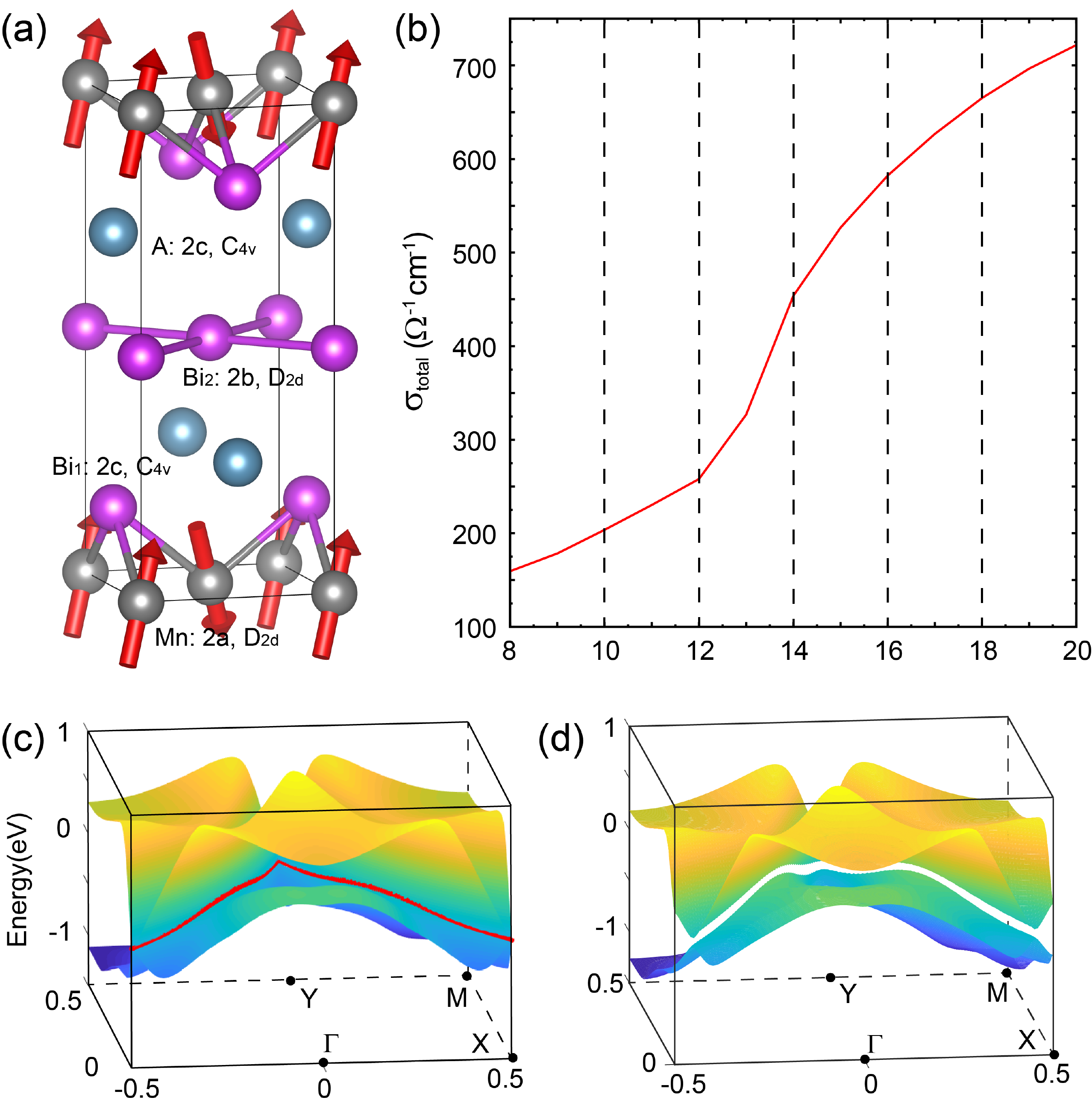}}
\caption{(color online) (a) The crystal structure and C-type AFM order of AMnBi$_2$ with spin canting  along the (1,1,0) direction. Wyckoff positions and corresponding site symmetry group are given. (b) Canting angle dependence of the AHC $\sigma_{AHC}$ in the CaMnBi$_2$. (c) and (d) The band structures in the $k_z$=0 plane without and with SOC, where red line is a nodal line protected by glide plane symmetry and SOC can break the nodal line into opened gaps.
\label{fig1} }
\end{figure}

 The crystal structure of AMnBi$_2$(A=Ca and Yb) with nonsymmorphic space group G=P4/nmm (No.129) is shown in Fig.\ref{fig1}(a), where the MnBi$_1$ layers possess an anti-PbO-type atom arrangement and there are a square lattice sheet of Bi$_2$ atoms. Mn and Bi$_2$ atoms occupy Wyckoff positions 2a \{(0,0,0), (1/2,1/2,0)\} and 2b \{(0,0,1/2), (1/2,1/2,1/2)\} respectively, which have the same corresponding site symmetry group $D_{2d}$,  while A and Bi$_1$ atoms occupy the same Wyckoff positions 2c \{(1/2,0,z), (0,1/2,-z)\} with corresponding site symmetry group $C_{4v}$. In the nonsymmorphic space group G=P4/nmm, the quotient group G/L is specified by 16 symmetry operations, where we denote L as the translation group with respect to the unit cell. The nonsymmorphic symmetry operations are $\tilde{C}_{2x}=\{C_{2x}|1/2, 1/2, 0\}$, $\tilde{C}_{2y}=\{C_{2y}|1/2, 1/2, 0\}$ and $\tilde{M}_{xy}=\{M_{xy}|1/2, 1/2, 0\}$ with respect to the original point at Mn site.

~~~Since the AMnBi$_2$ has AFM order at Mn atoms with spins along
the z axis, the following will analyze magnetic space groups and point groups.
We choose Mn site as the origin, and the 16 symmetry operations operations can
be specified equivalently as $D_{2 d} \oplus \{I|1/2,1/2,0\} D_{2 d}$, where $I$
is a space inversion and the generators of $D_{2 d}$ are $C_{2,z}$, $C_{2,xy}$
and $M_{y}$. Firstly, Due to spin at Mn atoms along the z axis, $C_{2,z}$ cannot
change the direction of spin while $C_{2,xy}$ and $M_{y}$ flip spin, indicating
that $C_{2,xy}$ and $M_{y}$ are broken by AFM order at Mn atoms. Combined with
time reversal symmetry $T$, we can get the magnetic point group
$\bar{4}2^{\prime}m^{\prime}$ corresponding to point group D$_{2d}$,
where the generators are $C_{2,z}$, $TC_{2,xy}$ and $TM_{y}$. Secondly,
Since the crystal symmetries in the $\tilde{I}D_{2 d}$ with
$\tilde{I}=\{I|1/2,1/2,0\}$ exchange Mn sublattice and inversion symmetry
$I$ can not flip spin, $C_{2,xy}$ and $M_{y}$ will be maintained while
$C_{2,z}$ is broken, suggesting that the magnetic generators corresponding
to $\tilde{I}D_{2 d}$  are $T\tilde{I}$, $T\tilde{I}C_{2,z}$,
$\tilde{I}C_{2,xy}$ and $\tilde{I}M_{y}$. Finally, based on the above
analysis, we find that the magnetic space group of AMnBi$_2$ with AFM
order along the z axis is $P4^{\prime}/n^{\prime}m^{\prime}m$ with
magnetic point group $4^{\prime}/m^{\prime}m^{\prime}m$. Due to $T\tilde{I}$ symmetry, the integration of Berry curvatures on BZ are forced to zero.

~~~Since a spin canting exists in the AMnBi$_2$ with
an in-plane ferromagnetic, and the magnitude of canting angle can be experimentally adjusted by doping\cite{Yang2020}. We take the same canting direction (1,1,0) as the papers\cite{Sergey2019,Yang2020}, which can break $T\tilde{I}$ symmetry, indicating that AHE can exist.
Then, the corresponding magnetic space group becomes $m^{\prime}m2^{\prime}$,
where the symmetries are TC$_{2,\bar{x}y}$,~M$_{11}$,~TM$_{z}$, E, which can host non-zero AHE.

~~~The AHC tensor can be written as

\begin{eqnarray}
\mathbf{\sigma}^{\gamma}_{\alpha \beta}&=&-\sum_{n}\frac{e^{2}}{\hbar} \int_{\mathrm{BZ}} \frac{d \mathbf{k}}{(2 \pi)^{3}} f_{n}(\mathbf{k}) \mathbf{\Omega}^{\gamma}_{n, \alpha \beta}(\mathbf{k}),\nonumber
\\
\mathbf{\Omega}^{\gamma}_{n, \alpha \beta}(\mathbf{k})&=&-2\operatorname{Im}\langle \nabla_{\alpha}u_n (\mathbf{k})| \nabla_{\beta}u_n (\mathbf{k})\rangle,
\end{eqnarray}

Under crystal symmetry $g$, the relation of $\mathbf{\Omega}^{\gamma}_{n, \alpha \beta}(\mathbf{k})$ between $\mathbf{k}$ and $g\mathbf{k}$ is given by

\begin{eqnarray}
\mathbf{\Omega}^{\gamma}_{n, \alpha \beta}(g\mathbf{k})&=&-2 \operatorname{Im}\langle\frac{\partial u_n (g\mathbf{k})}{\partial k_{\alpha}} | \frac{\partial u_n (g\mathbf{k})}{\partial k_{\beta}}\rangle,\nonumber
\\
&=&-2\sum_{\alpha^{\prime}\beta^{\prime}} \operatorname{Im}\langle\frac{\partial(g\mathbf{k})_{\alpha^{\prime}}}{\partial \mathbf{k}_{\alpha}}\frac{\partial u_n (g\mathbf{k})}{\partial (g\mathbf{k})_{\alpha^{\prime}}} | \frac{\partial(g\mathbf{k})_{\beta^{\prime}}}{\partial \mathbf{k}_{\beta}}\frac{\partial u_n (g\mathbf{k})}{\partial (g\mathbf{k})_{\beta^{\prime}}}\rangle,\nonumber
\\
&=&\sum_{\alpha^{\prime}\beta^{\prime}}\frac{\partial(g\mathbf{k})_{\alpha^{\prime}}}{\partial \mathbf{k}_{\alpha}}\frac{\partial(g\mathbf{k})_{\beta^{\prime}}}{\partial \mathbf{k}_{\beta}}\mathbf{\Omega}^{\gamma^{\prime}}_{n, \alpha^{\prime} \beta^{\prime}}(\mathbf{k}),\nonumber
\\
\mathbf{\Omega}^{\gamma^{\prime}}_{n, \alpha^{\prime} \beta^{\prime}}(\mathbf{k})&=&-2 \operatorname{Im}\langle\frac{\partial u_n (\mathbf{k})}{\partial \mathbf{k}_{\alpha^{\prime}}} | \frac{\partial u_n (\mathbf{k})}{\partial \mathbf{k}_{\beta^{\prime}}}\rangle,
\end{eqnarray}

Where $\alpha^{\prime}$ can not be equal to $\beta^{\prime}$. It is clear that the AHC is invariant under inversion symmetry $I$, namely $\mathbf{\Omega}^{\gamma}_{n,\alpha\beta}(\mathbf{k})=\mathbf{\Omega}^{\gamma}_{n,\alpha\beta}(I\mathbf{k})$. Then, we consider the constrain of the magnetic space group $m^{\prime}m2^{\prime}$ on AHC, where considering mirror symmetry $M_{11}(-\mathbf{k}_y,-\mathbf{k}_x,\mathbf{k}_z)$ is enough. Under mirror symmetry $M_{11}$, the transformation of Berry curvature $\mathbf{\Omega}^{\gamma}_{n, \alpha \beta}(\mathbf{k})$ can be given by
\begin{eqnarray}
\Omega_{xy}^{z}(\mathbf{k})=-\Omega_{xy}^{z}(M_{11}\mathbf{k}),~\Omega_{xz}^{y}(\mathbf{k})=-\Omega_{yz}^{x}(M_{11}\mathbf{k}),
\end{eqnarray}

Hence, under the constrain of magnetic space group $m^{\prime}m2^{\prime}$, the shape of the AHC tensors is
\begin{eqnarray}
\left(\begin{array}{ccc}
 0 & 0 & \sigma_{xz}\\
 0& 0& -\sigma_{xz}\\
 -\sigma_{xz}&\sigma_{xz} & 0
\end{array}\right),
\end{eqnarray}

\begin{figure*}
\centerline{\includegraphics[width=0.8\textwidth]{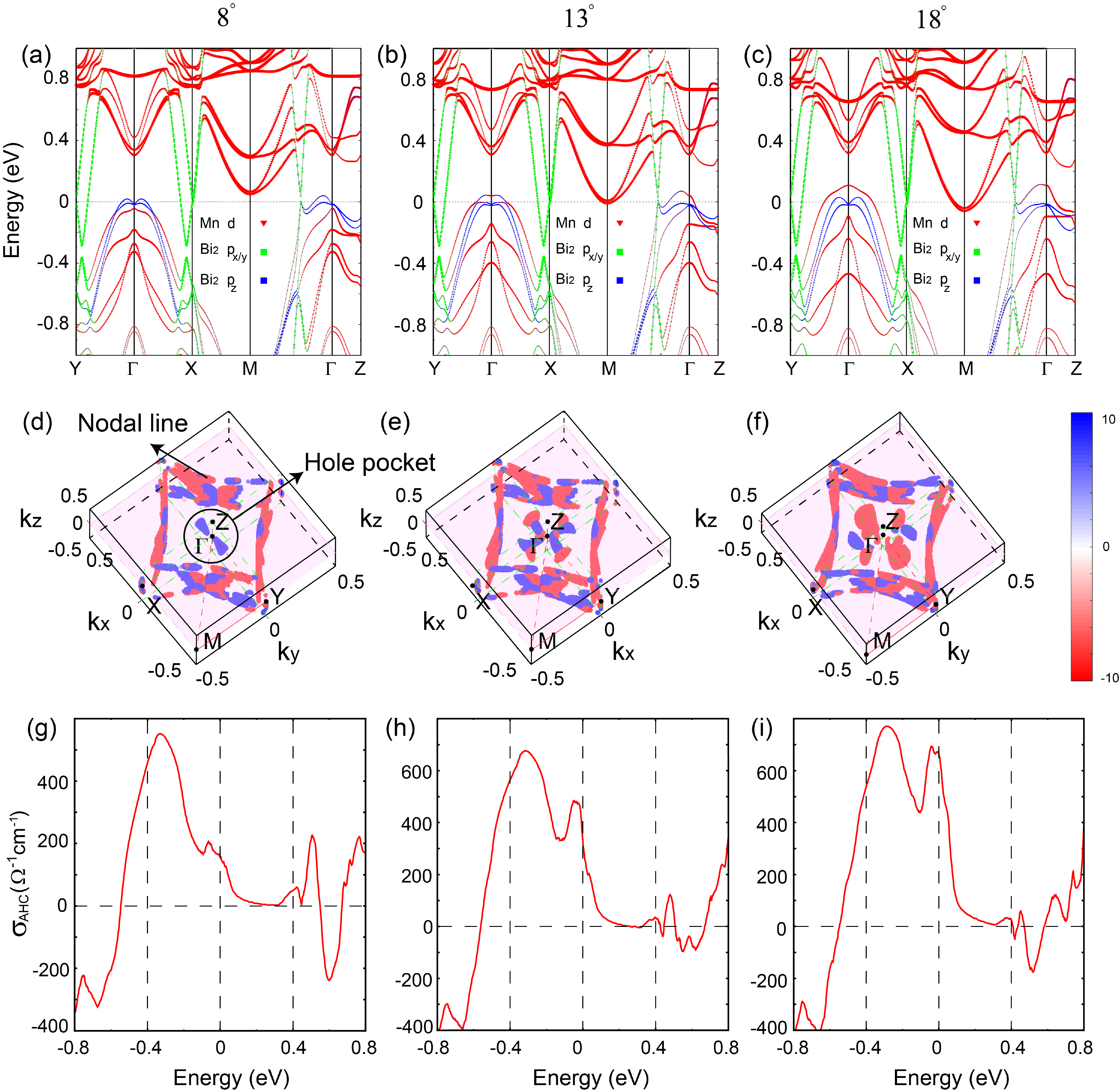}}
\caption{(color online) (a)-(c) The band structures with spin canted C-type AFM states at 8$^{\circ}$, 13$^{\circ}$ and 18$^{\circ}$ canting angles, respectively. The orbital characters of bands are represented by different colors. (d)-(f) show the local momentum distribution of Berry curvature in the BZ $\sigma_{total}$ at 8$^{\circ}$, 13$^{\circ}$ and 18$^{\circ}$ canting angles, respectively. (g)-(i) The energy-dependent AHC $\sigma_{AHC}$ at 8$^{\circ}$, 13$^{\circ}$ and 18$^{\circ}$ canting angles.
\label{fig2} }
\end{figure*}

\section{Ab initio analysis of anomalous transports}\label{S3}

Fig.\ref{fig1}(c) and (d) show band structures without and with SOC in the $k_z$=0 plane, where the half of BZ($-0.5<k_x<0.5$, and $0<k_y<0.5$) is shown for the convenience of seeing nodal line. There is a band inversion from intercalated Bi$_2$ $p_{x/y}$ orbitals, which lead to quasi two-dimensional band crossing. When the SOC is ignored, the crossing bands can be referred to as $K$ and $K+Q$ bands. Similar to the symmetry analysis of iron superconductors\cite{Wu2016}, the $K$ and $K+Q$ bands are the different eigenvalues of the glide plane symmetry\cite{Qiu2019}, indicating that the nodal ring from Bi$_2$ $p_{x/y}$ band crossing is robust, shown in Fig.\ref{fig1}(c). When the on-site SOC is included, the spin-flip term $\langle p_{x}, \sigma|H_{\mathrm{soc}}| p_{y}, \bar{\sigma}\rangle$ with the spin index $\sigma$ can be non-zero\cite{Qiu2019}. Hence, the crossing bands can hybridize each other and the nodal line can be gapped, shown in Fig.\ref{fig1}(d). In the following, we will calculate AHC and consider how does the nodal line contribute to AHC.

We calculated the AHC tensor of CaMnBi$_2$ with Fermi level lying at the charge neutral points, and Fig.\ref{fig1}(b) shows canting angle dependence of the AHC $\sigma_{AHC}$ with spin canting  along the (1,1,0) direction, where $\sigma_{AHC}=\sqrt{\sigma^2_{xz}+\sigma^2_{yz}}=\sqrt{2}\sigma_{xz}$ due to $\sigma_{yz}=-\sigma_{xz}$. At 8$^{\circ}$ canting angle, the AHC $\sigma_{AHC}$ has a small value of 159.50 $\Omega^{-1}$cm$^{-1}$, whereas it is a large value of 721.90 $\Omega^{-1}$cm$^{-1}$ at 20$^{\circ}$ canting angle. With the canting angle increasing from 8$^{\circ}$ to 20$^{\circ}$, the AHC $\sigma_{AHC}$ always keep growing, and is not saturated even if the canting angle reaches 20$^{\circ}$. What's more, between 12$^{\circ}$ and 14$^{\circ}$, the rate of AHC increasing is larger. To explain the variation of AHC with canting angle, we take three (8$^{\circ}$, 13$^{\circ}$ and 18$^{\circ}$) canting angles as examples to calculate band structures and local momentum distribution of Berry curvature in the BZ with spin canted C-type AFM states, where the magnitude of magnetic moment in the xy-plane (z-direction) are 0.53$\mu_B$ (3.81$\mu_B$), 0.99$\mu_B$ (3.74$\mu_B$) and 1.31$\mu_B$ (3.63$\mu_B$) respectively.

Firstly, Fig.\ref{fig2} (a) shows band structures at 8$^{\circ}$ canting angle. Similar to band structures without canting angle\cite{Yang2020}, near the Fermi level, the valence and conduction bands are mainly attributed to the p orbitals of the intercalated Bi$_2$ atoms. Because of ferromagnetic in the xy plane, the bands are split into spin-up and spin-down. The small hole pockets around the $\Gamma$ point are dominated by the intercalated Bi$_2$-p$_z$ (blue) orbital, which is remarkably split due to hybridizing with Mn d orbitals; The electronic pockets from nodal line mainly originate from the intercalated Bi$_2$-p$_{x/y}$ (green) orbitals, which is less sensitive to the ferromagnetic in the xy plane and is hardly split. Fig.\ref{fig2} (d) shows the local momentum distribution of Berry curvature in the BZ at 8$^{\circ}$ canting angle. Since the value of AHC $\sigma_{AHC}$ is 159.50 $\Omega^{-1}$cm$^{-1}$ at 8$^{\circ}$ canting angle, the total Berry curvature contributed by the hole pockets near $\Gamma$ is positive and small while the total Berry curvature from electronic pockets is negative and large. The nodal line has dispersion in the BZ, which can result in electronic pockets at the fermi level, and hence the nodal line is mainly responsible for non-zero AHC.

Secondly, when the canting angle is increased to 13$^{\circ}$, the split of band structures become more larger as the magnitude of ferromagnetic in the xy-plane increases. Then, the electronic pockets from nodal line have more splits, which maybe strengthen AHC; The Mn d-orbital bands can also attribute a hole pocket near $\Gamma$ point which can result in a positive Berry curvature, shown in Fig.\ref{fig2} (b) and (e). Hence, the AHC of 13$^{\circ}$ canting angle becomes more greater than that of 8$^{\circ}$, shown in Fig.\ref{fig1}(b). Finally, when the canting angle is further increased to 18$^{\circ}$, a big hole pocket near $\Gamma$ point is completely contributed by the Mn-$d$ orbitals due to large band splits induced by strong ferromagnetism. Then, the big hole pocket can result in a big and positive Berry curvature, indicating that the AHC of 18$^{\circ}$ canting angle becomes more larger, shown in Fig.\ref{fig2} (c) and (f).

Here, summarizing the above analysis. We find that the ferromagnetism in the xy plane becomes larger with the increasing of the canting angle, which can lead to greater splits of band structures. Then, the hole pocket contributed by Mn d orbitals will become larger, and can result in big and negative Berry curvature. Moreover, the electronic pockets from nodal line mainly contribute to AHC, indicating that the nodal line is responsible for non-zero AHC. As the ferromagnetism increases, the electronic pockets have more splits, which can strengthen AHC. Therefore, the AHC always keeps growing as the canting angle increases.

\begin{figure}
\centerline{\includegraphics[width=0.5\textwidth]{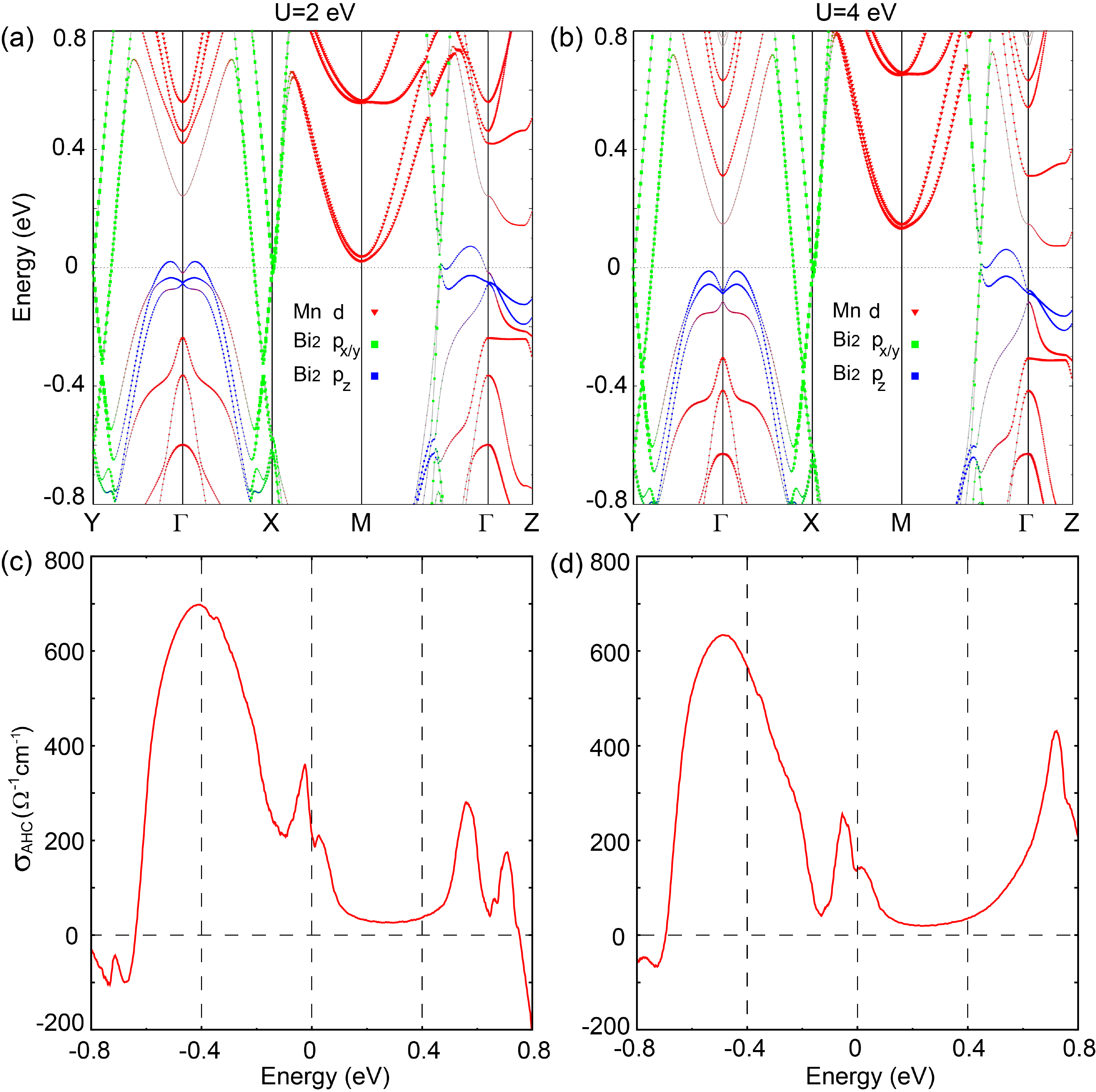}}
\caption{(color online) The band structures and corresponding energy-dependent AHC tensor $\sigma_{AHC}$ with the effective on-site Coulomb U being (a,c) 2 eV and  (b,d) 4 eV for Mn d orbitals at 18$^{\circ}$ canting angles in the GGA+U calculation.
\label{fig5} }
\end{figure}

~~~To further explore the effect of holes and electrons doping on AHC $\sigma_{AHC}$, Fig.\ref{fig2}(d)-(f) show the energy-dependent AHC $\sigma_{AHC}$ calculated from the Berry curvature at 8$^{\circ}$, 13$^{\circ}$ and 18$^{\circ}$ canting angles. When the Fermi level is at the charge neutral points, the AHC $\sigma_{AHC}$ are 110, 230 and 470 $\Omega^{-1}$cm$^{-1}$ respectively, indicating that the AHC keep growing as canting angle increases. At 13$^{\circ}$ and 18$^{\circ}$ canting angles, a peak in $\sigma_{AHC}$ appears around Fermi level, shown in Fig.\ref{fig2}(e) and (f). Additionally, it is clear that doping electrons can reduce AHC while doping holes strengthen AHC.

~~~In order to consider the effect of correlation effect on AHC in the canted C-type AFM order, Fig.\ref{fig5} shows band structures and AHC with the effective on-site Coulomb U being 2 eV and 4 eV for Mn d orbitals at 18$^{\circ}$ canting angles in the GGA+U calculation. The gap between Mn d orbitals (red) increases as U increases, while band structures attributed by intercalated Bi$_2$ p orbitals (green and blue) remain unchanged, shown in Fig.\ref{fig5}(a) and (b). Compared with band structures without U, the d-orbital bands with U stay away from Fermi level, and the bands at Fermi level are mainly attributed to the Bi$_2$ p orbitals. Fig.\ref{fig5}(c) and (d) show energy-dependent AHC tensor $\sigma_{AHC}$ with U being 2 eV and 4 eV for Mn d orbitals at 18$^{\circ}$ canting angles, and AHC tensors $\sigma_{AHC}$ with U are small compared with that without U, indicating that the correlation effect can suppress AHC.

~~~Since the spin canting exist in the YbMnBi$_2$, we calculate AHC tensors and band structures of YbMnBi$_2$ at different canting angles. Similar to CaMnBi$_2$, the ferromagnetism in the xy plane becomes larger with the increasing of the canting angle, and the hole pocket at $\Gamma$ point contributed by Mn d orbitals will become larger, which can result in big and negative Berry curvature. Therefore, the AHC $\sigma_{AHC}$ in the YbMnBi$_2$ also increases as the canting angle increases. The energy-dependent AHC $\sigma_{AHC}$ is also calculated at different canting angles, and we find that doping electrons can suppress AHC while doping holes strengthen AHC.

\section{Conclusion}\label{S4}

~~~In summary, the AHE in the layered ternary materials AMnBi$_2$(A=Ca and Yb) with spin canted C-type antiferromagnetic order can be studied based on first-principles calculations and symmetry analysis. By taking the canting along the (1,1,0) direction, the magnetic space group is $m^{\prime}m2^{\prime}$, and only $\sigma_{xz/yz}$ in the AHC tensors are non-zero due to mirror symmetry. We find that the AHC $\sigma_{AHC}$ always keep growing with the canting angle increasing, and doping electrons can reduce AHC while doping holes strengthen AHC. The correlation effect from Mn $d$ orbitals is considered, and can suppress AHC. This work provides an effective design strategy and corresponding materials to have strong AHE.

\section{ACKNOWLEDGMENTS}

We thank Yu Pan for helpful discussions. This work was financially
supported by the ERC Advanced Grant No. 291472 'Idea Heusler',
ERC Advanced Grant No. 742068 'TOPMAT', SKYTOP with grant No. 824123, ASPIN with grant No. 766566, DAAD grant No. 57559136. Some of
our calculations were carried out on the Cobra cluster of
MPCDF, Max Planck society.

\end{document}